\documentclass[11pt]{article}%
\usepackage{amsmath}
\usepackage{moriond,epsfig}
\usepackage{graphicx}%
\usepackage{amsfonts}%
\usepackage{amssymb}
\def\be{\begin{equation}}
\def\ee{\end{equation}}
\def\bea{\begin{eqnarray}}
\def\eea{\end{eqnarray}}

\begin{document}

\vspace*{4cm}
\title{LONG-DISTANCE EFFECTS IN RARE K DECAYS}
\author{ C. SMITH }
\address{Theory Group, Laboratori Nazionali di Frascati, Via E. Fermi, 40, I-00044 Frascati, Italy.}
\maketitle
\abstracts{Rare K decays provide for very clean tests of the Standard Model, and are
especially suited to search for New Physics signal. In this talk, recent progresses in
the estimation of long-distance effects induced by light quarks in $K_L\rightarrow
\pi^0\mu^+\mu^-$ and $K^+\rightarrow\pi^+\nu\bar\nu$ are reported.}

\section{Introduction}

The rare $K$ decays considered in this talk proceed through Flavor Changing
Neutral Currents, arising at loop-level in the Standard Model. What makes them
specially interesting is that short-distance (SD) effects contribute
significantly to their decay rates, and that long-distance (LD) hadronic
effects are under theoretical control. They are thus ideal to constrain the
Standard Model by precise extraction of CKM parameters. In addition, being
suppressed in the SM and being driven by SD physics give them good sensitivity
to possible New Physics, complementary to direct searches. The SM theoretical
predictions are%
\begin{equation}%
\begin{array}
[c]{ll}%
\mathcal{B}\left(  K_{L}\rightarrow\pi^{0}\nu\bar{\nu}\right)  =\left(
3.0\pm0.6\right)  \times10^{-11}\;\;\; & \mathcal{B}\left(  K_{L}%
\rightarrow\pi^{0}e^{+}e^{-}\right)  =3.7_{-0.9}^{+1.1}\times10^{-11}\\
\mathcal{B}\left(  K^{+}\rightarrow\pi^{+}\nu\bar{\nu}\right)  =\left(
7.8\pm1.2\right)  \times10^{-11}\;\;\;\; & \mathcal{B}\left(  K_{L}%
\rightarrow\pi^{0}\mu^{+}\mu^{-}\right)  =\left(  1.5\pm0.3\right)
\times10^{-11}%
\end{array}
\label{Eq0}%
\end{equation}
Experimentally, KTeV\thinspace\cite{KTeV} has set upper limits for the neutral
modes, and AGS-E787/E979\thinspace\cite{E787E949} found three events for the
charged one.

The leading parts of the effective Hamiltonians relevant to the study of these
modes are\thinspace\cite{BLMM}%
\begin{gather}
H_{eff}\left(  \bar{s}d\rightarrow\nu\bar{\nu}\right)  \sim\frac{G_{F}}%
{\sqrt{2}}\left(  y_{\nu}(\bar{s}d)_{V-A}(\nu\bar{\nu})_{V-A}+h.c.\right)
\label{Eq1}\\
H_{eff}\left(  \bar{s}d\rightarrow\ell^{+}\ell^{-}\right)  \sim\frac{G_{F}%
}{\sqrt{2}}\left(  y_{7V}(\bar{s}d)_{V-A}(\ell^{+}\ell^{-})_{V}+y_{7A}(\bar
{s}d)_{V-A}(\ell^{+}\ell^{-})_{A}+h.c.\right)  \label{Eq2}\\
y_{\nu}=\frac{\alpha}{2\pi}\sum\lambda_{q}\frac{X_{0}\left(  x_{q}\right)
}{\sin^{2}\theta_{W}},\;y_{7A}=\frac{\alpha}{2\pi}\sum\lambda_{q}%
\frac{-Y_{0}\left(  x_{q}\right)  }{\sin^{2}\theta_{W}},\;y_{7V}=\frac{\alpha
}{2\pi}\sum\lambda_{q}\left[  \frac{Y_{0}\left(  x_{q}\right)  }{\sin
^{2}\theta_{W}}-4Z_{0}\left(  x_{q}\right)  \right]  \label{Eq3}%
\end{gather}
with $\lambda_{q}=V_{qs}^{\ast}V_{qd}$, $x_{q}=m_{q}^{2}/M_{W}^{2}$, summation
over $q=u,c,t$ and the Inami-Lim functions $X_{0}=C_{0}^{Z}-4B_{0}^{W}$,
$Y_{0}=C_{0}^{Z}-B_{0}^{W}$ and $Z_{0}=C_{0}^{Z}+D_{0}^{\gamma}/4$ depicted in
Fig.\thinspace1.%
\[
\text{%
{\parbox[b]{4.3111in}{\begin{center}
\includegraphics[
height=0.8181in,
width=4.3111in
]%
{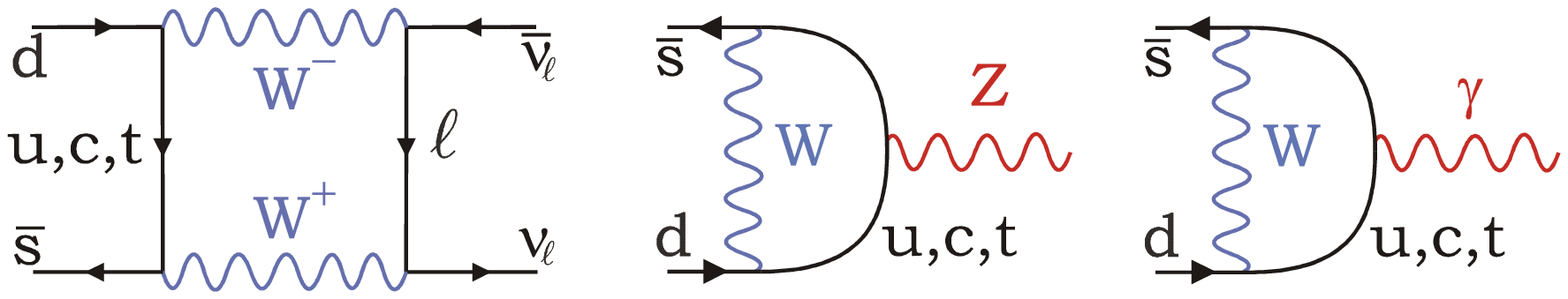}%
\\
Figure 1: The $W^{\pm}$ box, Z penguin and photon penguin diagrams.
\end{center}}}%
}%
\]

In $H_{eff}$, the operators are new interactions, while the Wilson
coefficients $y_{i}$, encoding the effects of heavy particles, their coupling
constants. Combining the known scaling of CKM matrix elements with the
behavior of the Inami-Lim functions as functions of quark masses, one can
assess of the relative strengths of the $u,c,t$ contributions, and thereby of
the SD or LD nature of the process. For example, for the simplest and cleanest
rare decay mode, $K_{L}\rightarrow\pi^{0}\nu\bar{\nu}$, taking the matrix
element of Eq.\thinspace\ref{Eq1},
\begin{equation}
\mathcal{M}\left(  K_{L}\rightarrow\pi^{0}\nu\bar{\nu}\right)  =\frac{G_{F}%
}{\sqrt{2}}\left\langle \pi^{0}\right|  H_{eff}\left|  K_{L}\right\rangle
(\nu\bar{\nu})_{V-A}=G_{F}\operatorname{Im}y_{\nu}\left\langle \pi^{0}\right|
(\bar{s}d)_{V}\left|  K^{0}\right\rangle (\nu\bar{\nu})_{V-A}\label{Eq4}%
\end{equation}
The CKM structure and GIM hard breaking ($X_{0}\left(  x\right)  \sim x$)
suppress all light-quark effects, leaving only the top quark contribution:%
\begin{equation}%
\begin{array}
[c]{cccccccccc}%
\operatorname{Im}y_{\nu}\sim & \underbrace{\operatorname{Im}\lambda_{u}} &
X_{0}\left(  x_{u}\right)   & + & \underbrace{\operatorname{Im}\lambda_{c}} &
\underbrace{X_{0}\left(  x_{c}\right)  } & + & \underbrace{\operatorname{Im}%
\lambda_{t}} & \underbrace{X_{0}\left(  x_{t}\right)  } & \approx
\operatorname{Im}\lambda_{t}X_{0}\left(  x_{t}\right)  \\
& =0 &  &  & \mathcal{O}(\lambda^{5}) & \mathcal{O}(10^{-4}) &  &
\mathcal{O}(\lambda^{5}) & \mathcal{O}(1) &
\end{array}
\label{Eq5}%
\end{equation}
A second important fact is that the matrix element $\langle\pi^{0}|(\bar
{s}d)_{V}|K^{0}\rangle$ can be extracted from the well-measured decay
$K^{+}\rightarrow\pi^{0}\ell^{+}\nu_{\ell}$ using isospin symmetry, hence
hadronic uncertainties are small. From these properties, $\mathcal{B}\left(
K_{L}\rightarrow\pi^{0}\nu\bar{\nu}\right)  \overset{SM}{=}2.6\cdot
10^{-11}\overset{QCD\,}{\rightarrow}\left(  3.0\pm0.6\right)  \cdot10^{-11}$,
with the error coming mostly from the uncertainty on the CKM element\thinspace
\cite{BLMM}.

\section{CP-conserving contributions to $K_{L}\rightarrow\pi^{0}\ell^{+}%
\ell^{-}$}

Modes involving a charged lepton pair are more complicated as they
interact with photons. Three types of processes are relevant: the direct
CP-violating (DCPV), indirect CP-violating (ICPV) and CP-conserving (CPC) one, see Fig.\thinspace2. 
While the theoretical complexity increases for
$\ell^{+}\ell^{-}$ modes, recent experimental results for $K_{S}\rightarrow
\pi^{0}\ell^{+}\ell^{-}$ \cite{NA48as} and $K_{L}\rightarrow\pi^{0}%
\gamma\gamma$ \cite{KTeV_KLpgg,NA48_KLpgg} permit reliable theoretical
estimates for ICPV and CPC. The relative sizes of the three contributions are%
\begin{equation}
\text{ }%
\begin{tabular}
[c]{|lllll|}\hline
& DCPV & ICPV & CPC-2$^{++}$ & CPC-0$^{++}$\\\hline
$K_{L}\rightarrow\pi^{0}\nu\bar{\nu}$ & $100\%$ & $\left(  \sim1\%\right)  $ &
$-$ & $-$\\
$K_{L}\rightarrow\pi^{0}e^{+}e^{-}$ & $40\%$ & $60\%$ & $\left(  <3\%\right)
$ & $-$\\
$K_{L}\rightarrow\pi^{0}\mu^{+}\mu^{-}$ & $30\%$ & $35\%$ & $-$ &
$35\%$\\\hline
\end{tabular}
\ \ \ \label{Eq6}%
\end{equation}
where $-$ means $<0.1\%$. The CPC contribution proceeds through two photons
which can be in a scalar $0^{++}$ (helicity suppressed for $e^{+}e^{-}$) or
tensor $2^{++}$ state (phase-space suppressed, especially for $\mu^{+}\mu^{-}%
$). Our work was to estimate the $0^{++}$ CPC contribution\thinspace
\cite{OurWorkMu}.%
\[
\text{%
\raisebox{-0.0095in}{\parbox[b]{5.3402in}{\begin{center}
\includegraphics[
height=0.8588in,
width=5.3402in
]%
{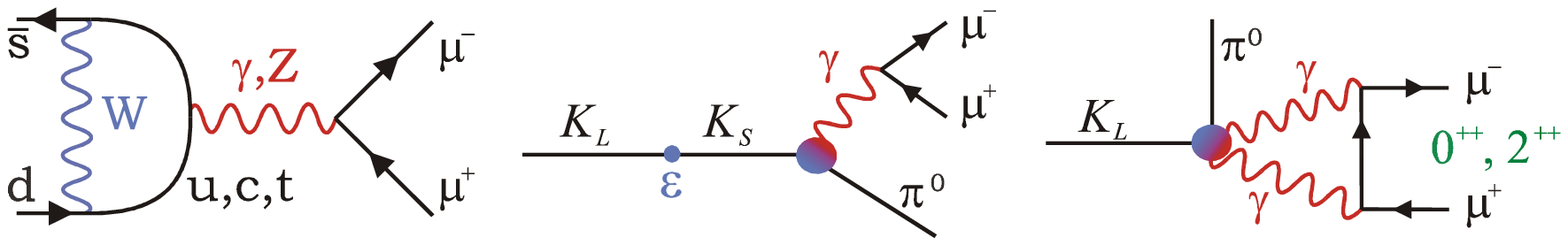}%
\\
Figure 2: Direct CPV, Indirect CPV and CPC contributions.
\end{center}}}%
}%
\]

\paragraph{CPV contributions:}

The photon penguin now plays a role. Since it behaves as $D_{0}^{\gamma
}\left(  x\right)  \rightarrow\ln x$ when $x\rightarrow0$, light quarks may
contribute significantly. For direct CPV, one should again look at the scaling
of the $u,c,t$ contributions to the imaginary parts of $y_{7A,V}$%
\begin{align}
\operatorname{Im}y_{7A}  &  \sim\operatorname{Im}\lambda_{t}Y_{0}\left(
x_{t}\right)  \rightarrow-\left(  0.68\pm0.03\right)  \operatorname{Im}%
\lambda_{t}\times10^{-4}\nonumber\\
\operatorname{Im}y_{7V}  &  \sim\operatorname{Im}\lambda_{c}D_{0}^{\gamma
}\left(  x_{c}\right)  +\operatorname{Im}\lambda_{t}C_{0}^{Z}\left(
x_{t}\right)  \rightarrow\left(  0.73\pm0.04\right)  \operatorname{Im}%
\lambda_{t}\times10^{-4} \label{Eq7}%
\end{align}
with roughly equal $c$ and $t$-quark contributions for $y_{7V}$, and where NLO
QCD effects are included\thinspace\cite{BLMM}. In terms of these, the DCPV
contributions are\thinspace\cite{OurWorkMu,BDI}%
\begin{align}
\mathcal{B}\left(  K_{L}\rightarrow\pi^{0}e^{+}e^{-}\right)  _{DCPV}  &
=(2.67(\operatorname{Im}y_{7V}^{2}+\operatorname{Im}y_{7A}^{2})+\,\sim
0\;\operatorname{Im}y_{7A}^{2})\cdot10^{-12}\nonumber\\
\mathcal{B}\left(  K_{L}\rightarrow\pi^{0}\mu^{+}\mu^{-}\right)  _{DCVP}  &
=(0.63(\operatorname{Im}y_{7V}^{2}+\operatorname{Im}y_{7A}^{2}%
)+0.85\;\operatorname{Im}y_{7A}^{2})\cdot10^{-12} \label{Eq8}%
\end{align}
An extra helicity suppressed piece appears for the $\mu^{+}\mu^{-}$, giving
different sensitivities to the two modes on the SD physics, and therefore also
to possible New Physics.%
\[
\text{%
{\parbox[b]{4.6648in}{\begin{center}
\includegraphics[
height=1.7677in,
width=4.6648in
]%
{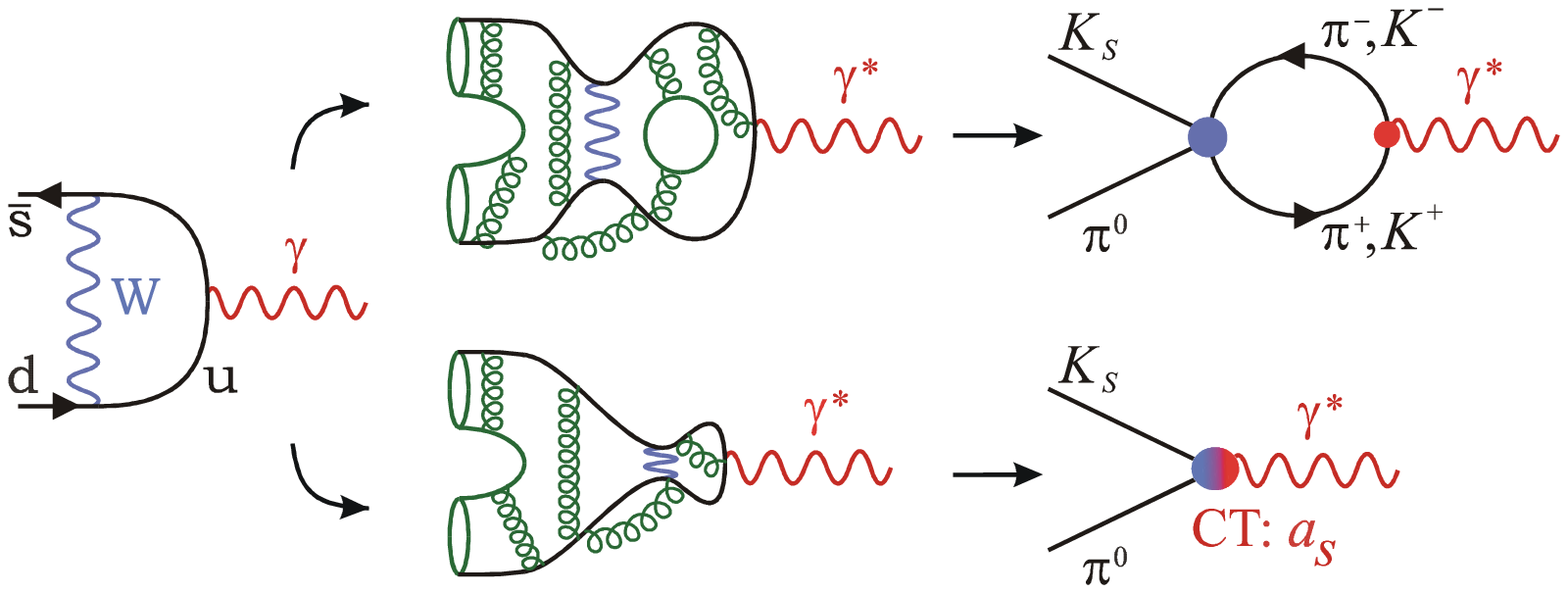}%
\\
Figure 3: Switching to mesons for the LD-dominated ICPV piece.
\end{center}}}%
}%
\]

For indirect CPV, $\mathcal{M}\left(  K_{L}\rightarrow\pi^{0}\ell^{+}\ell
^{-}\right)  =\varepsilon\mathcal{M}\left(  K_{S}\rightarrow\pi^{0}\ell
^{+}\ell^{-}\right)  $, one has to analyze the real part of the $y_{7A,V}$
(see Eq.\thinspace\ref{Eq4}). These are completely dominated by the $u$-quark
contribution, both from the CKM scaling and the GIM-breaking induced by
$D_{0}^{\gamma}\left(  x\right)  $ and $\operatorname{Re}y_{7V}\sim
\operatorname{Re}\lambda_{u}D_{0}^{\gamma}\left(  x_{u}\right)  \gg
\operatorname{Re}y_{7A}$. Being LD dominated, one has to switch to the meson
world, and the process is dealt with using Chiral Perturbation theory. It was
found in\thinspace\cite{DEIP} that loops have a small effect and both decays
are dominated by a common counterterm%
\begin{equation}
\mathcal{B}(K_{S}\rightarrow\pi^{0}e^{+}e^{-})=5.2a_{S}^{2}\cdot
10^{-9},\;\;\mathcal{B}(K_{S}\rightarrow\pi^{0}\mu^{+}\mu^{-})=1.2a_{S}%
^{2}\cdot10^{-9} \label{Eq9}%
\end{equation}
Recent NA48 measurements of both modes\thinspace\cite{NA48as} give $\left|
a_{S}\right|  =1.2\pm0.2$.

\paragraph{CPC contributions:}

The leading order is obtained from a $\pi^{\pm}$ or $K^{\pm}$ loop followed by
a $\gamma\gamma$ loop. This process can be factorized into a $K_{L}%
\rightarrow\pi^{0}P^{+}P^{-}$ ($P=\pi,K$) form factor convoluted with the
two-loop amplitude for $(P^{+}P^{-})_{0^{++}}\rightarrow\gamma\gamma
\rightarrow\ell^{+}\ell^{-}$, as long as the form factor depends on
$z\sim(p_{P^{+}}+p_{P^{-}})^{2}$ only (see Fig.\thinspace4). The crucial point
is that for a large range of parametrization of this dependence, the ratio
$R_{\gamma\gamma}=\mathcal{B}\left(  K_{L}\rightarrow\pi\ell^{+}\ell
^{-}\right)  /\mathcal{B}\left(  K_{L}\rightarrow\pi^{0}\gamma\gamma\right)  $
is stable even if both the $\ell^{+}\ell^{-}$ and $\gamma\gamma$ spectra and
rates vary much\thinspace\cite{OurWorkMu}. For dynamical reasons, the
$\ell^{+}\ell^{-}$ and $\gamma\gamma$ modes react similarly to modulations in
the distribution of momenta entering the scalar subprocess (i.e., to
$a_{1}\left(  z\right)  $ in Fig.\thinspace4). Given this observation, we
infer the branching ratios of $\ell^{+}\ell^{-}$ modes from the experimental
measurement of the $\gamma\gamma$ one. Some higher order chiral corrections
are thus included in our result, in particular the $\mathcal{O}\left(
p^{6}\right)  $ chiral counterterms (with their VMD contents) needed to
describe both the rate and spectrum for $K_{L}\rightarrow\pi^{0}\gamma\gamma$.
The stability of $R_{\gamma\gamma}$ is the key dynamical feature permitting
such an extrapolation, and thereby getting a reliable estimation for $\ell
^{+}\ell^{-}$ modes.

Numerically, taking $\mathcal{B}^{\exp}\left(  K_{L}\rightarrow\pi^{0}%
\gamma\gamma\right)  =\left(  1.42\pm0.13\right)  \cdot10^{-6}$ as the average
of KTeV\thinspace\cite{KTeV_KLpgg} and NA48\thinspace\cite{NA48_KLpgg}
measurements, we find $\mathcal{B}\left(  K_{L}\rightarrow\pi^{0}\mu^{+}%
\mu^{-}\right)  _{CPC}^{0^{++}}=\left(  5.2\pm1.6\right)  \cdot10^{-12}$, with
a conservative error estimate of 30\%. For the $e^{+}e^{-}$ mode, the BR is
$\mathcal{O}\left(  10^{-14}\right)  $, hence completely negligible.%
\[
\text{%
{\parbox[b]{3.691in}{\begin{center}
\includegraphics[
height=0.9132in,
width=3.691in
]%
{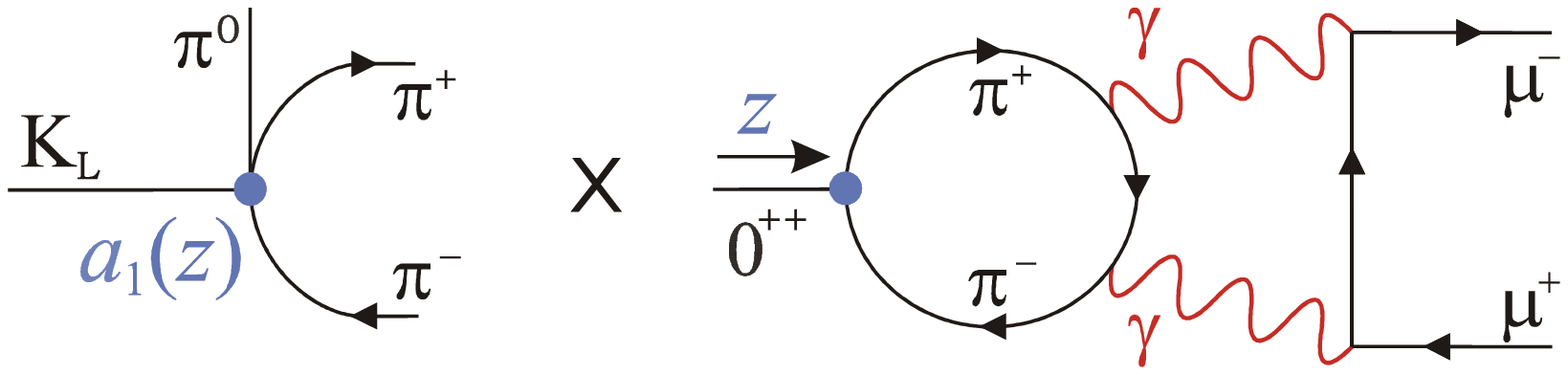}%
\\
Figure 4: Factorization of the CPC contribution.
\end{center}}}%
}%
\]

CPC-2$^{++}$ contributions were discussed in\thinspace\cite{BDI}. These
authors used strong constraints from the experimental low-energy part of the
photon spectrum in $K_{L}\rightarrow\pi^{0}\gamma\gamma$ to set the upper
limits $\mathcal{B}\left(  K_{L}\rightarrow\pi^{0}e^{+}e^{-}\right)
_{CPC}^{2^{++}}<3\cdot10^{-12}$ and $\mathcal{B}\left(  K_{L}\rightarrow
\pi^{0}\mu^{+}\mu^{-}\right)  _{CPC}^{2^{++}}<5\cdot10^{-14}$. These bounds
are very conservative, 2$^{++}$ contributions are probably smaller and can be neglected.

\paragraph{Complete SM prediction:}

The final parametrizations are, in the Standard Model%
\begin{align}
\mathcal{B}\left(  K_{L}\rightarrow\pi^{0}e^{+}e^{-}\right)   &
\approx(2.4\kappa^{2}\pm6.2\left|  a_{S}\right|  \kappa+15.7\left|
a_{S}\right|  ^{2})\times10^{-12}\nonumber\\
\mathcal{B}\left(  K_{L}\rightarrow\pi^{0}\mu^{+}\mu^{-}\right)   &
\approx(1.0\kappa^{2}\pm1.6\left|  a_{S}\right|  \kappa+3.7\left|
a_{S}\right|  ^{2}+5.2)\times10^{-12}\label{Eq10}%
\end{align}
with $\kappa=10^{4}\operatorname{Im}\lambda_{t}=1.36\pm0.12$. The interference
sign between DCPV and ICPV is not fixed by experiment, but two independent
theoretical analyses point toward a positive sign\thinspace\cite{BDI,FGD}.

The different sensitivities to SD physics can be illustrated by taking a
specific model\thinspace\cite{OurWorkMu}. Enhanced electroweak
penguins\thinspace\cite{BFRS} could lead to an enhancement of SD effects
$y_{7V}^{EEWP}=1.2\times y_{7V}^{SM}$, $y_{7A}^{EEWP}=4.7\times y_{7A}^{SM}$,
leading to the situation depicted in Fig.\thinspace5. A combined observation
of the two modes increases the sensitivity to New Physics, and in addition,
informations on its nature can be extracted: the difference in the vector and
axial vector currents manifests itself in a central value away from the
$\operatorname{Im}\lambda_{t}$ curve.%
\begin{gather*}
\text{%
{\includegraphics[
height=2.6308in,
width=3.3572in
]%
{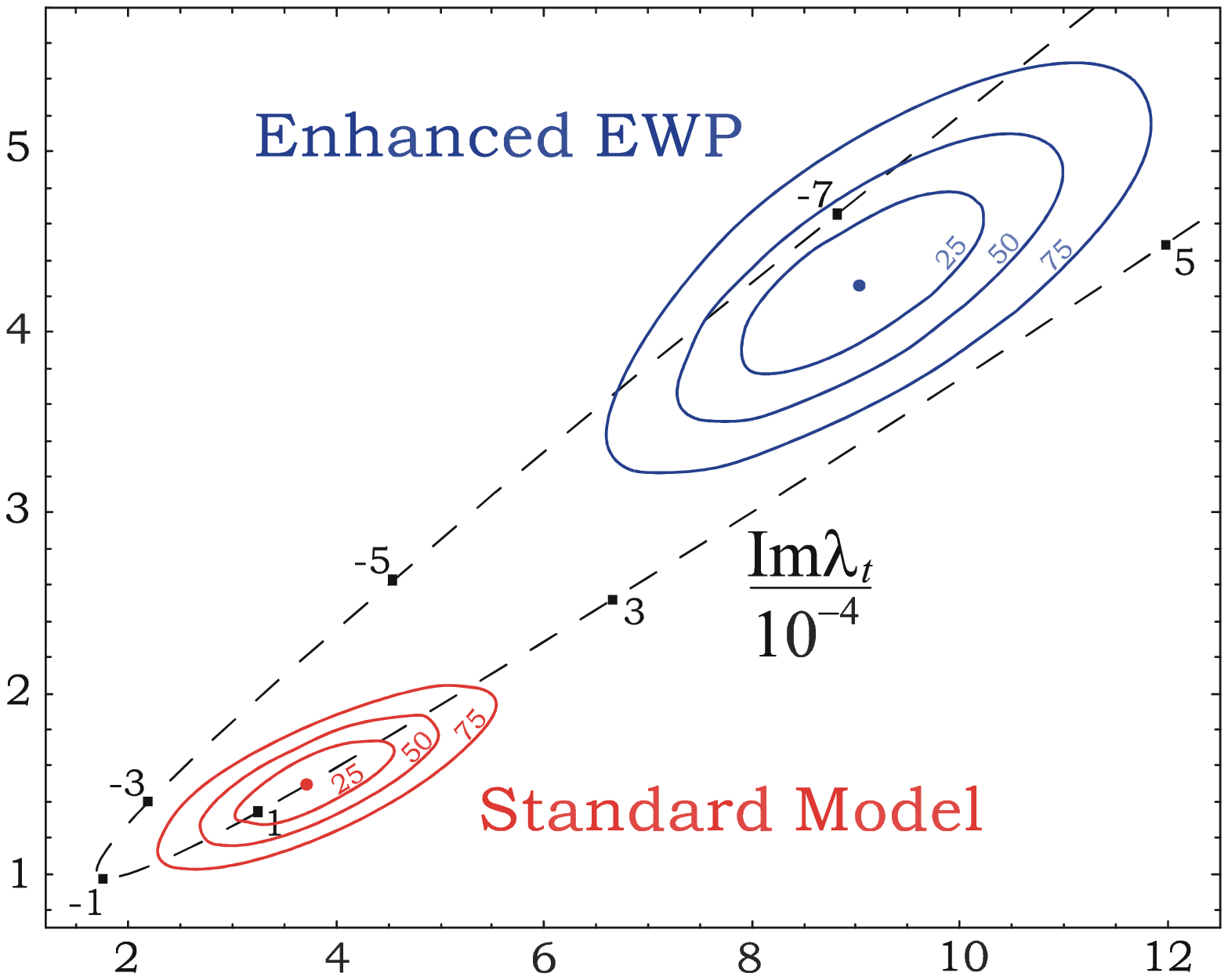}%
}%
}\\
\text{Figure 5: BR of the }\mu^{+}\mu^{-}\text{against the }e^{+}e^{-}\text{
mode (}\times10^{-11}\text{).}%
\end{gather*}

\section{Light-quarks in $K^{+}\rightarrow\pi^{+}\nu\bar{\nu}$}

The general structure is\thinspace\cite{BLMM}, with $\kappa^{+}\sim3\alpha
^{2}BR\left(  K^{+}\rightarrow\pi^{0}e^{+}\nu\right)  /2\pi^{2}\lambda^{2}%
\sin^{4}\theta_{W}$:%
\begin{equation}
\mathcal{B}\left(  K^{+}\rightarrow\pi^{+}\nu\bar{\nu}\right)  =\kappa
^{+}\left(  \left|  \operatorname{Im}\lambda_{t}X\left(  x_{t}\right)
\right|  ^{2}+\left|  \operatorname{Re}\lambda_{t}X\left(  x_{t}\right)
+\operatorname{Re}\lambda_{c}X\left(  x_{c}\right)  \right|  ^{2}\right)
\label{Eq11}%
\end{equation}
For the $\operatorname{Im}\lambda_{t}$ part, see Eq.\thinspace\ref{Eq5}. For
the real part, the CKM structure compensates the GIM suppression, and the $t$
and $c$ contributions are similar in size (68\% vs. 32\%). Let us analyze the
various uncertainties, with the purpose of precision physics in mind. The top
quark effects\thinspace\cite{BLMM} are known to within $3\%$, $X\left(
x_{t}\right)  \overset{NLO}{=}1.529\pm0.042$. The c-quark effects are also
known at NLO\thinspace\cite{BLMM}, but given the low scale set by $m_{c}$,
this corresponds to a $18\%$ error, $X\left(  x_{c}\right)  \overset{NLO}%
{=}\lambda^{4}\left(  0.39\pm0.07\right)  $.

Our goal was to analyze two subleading effects, on which control is needed to
get down to a few \% precision at the BR level\thinspace\cite{OurWorkNu}.
First there are the $c$-quark dimension eight operators\thinspace\cite{Falk},
for which a naive estimate of the possible impact on $X\left(  x_{c}\right)  $
would be $\mathcal{O}(m_{K}^{2}/m_{c}^{2}\sim15\%)$. Then, residual $u$-quark
effects\thinspace\cite{LD}, which are purely LD, could lead to $\mathcal{O}%
(\Lambda_{QCD}^{2}/m_{c}^{2}\sim10\%)$ corrections to $X\left(  x_{c}\right)
$. In the general OPE expansion, these effects are schematically depicted as%
\[%
\begin{tabular}
[c]{l}%
\raisebox{-0.3753in}{\includegraphics[
height=0.8726in,
width=3.243in
]%
{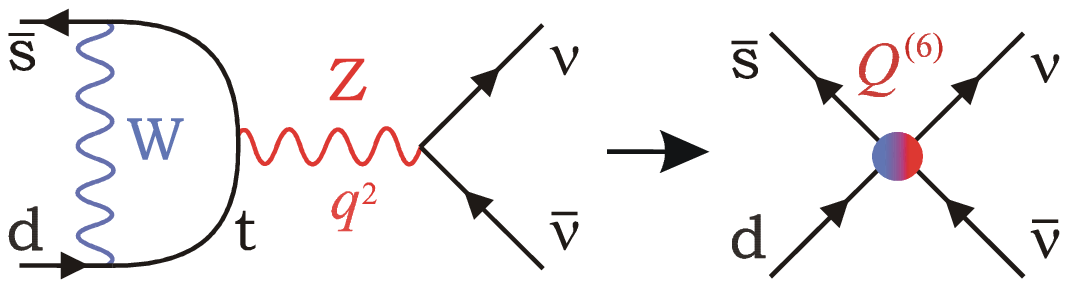}%
}%
\ \ \ \ \ \ \ \ \ \ \ \ \ \ \ \ \ \ \ \ \ \ \ \ \ \ \ \ \ $%
\begin{array}
[c]{c}%
\text{t-quark contribution }\\
\text{remains local}%
\end{array}
$\\%
\raisebox{-0.3425in}{\includegraphics[
height=0.8406in,
width=4.6328in
]%
{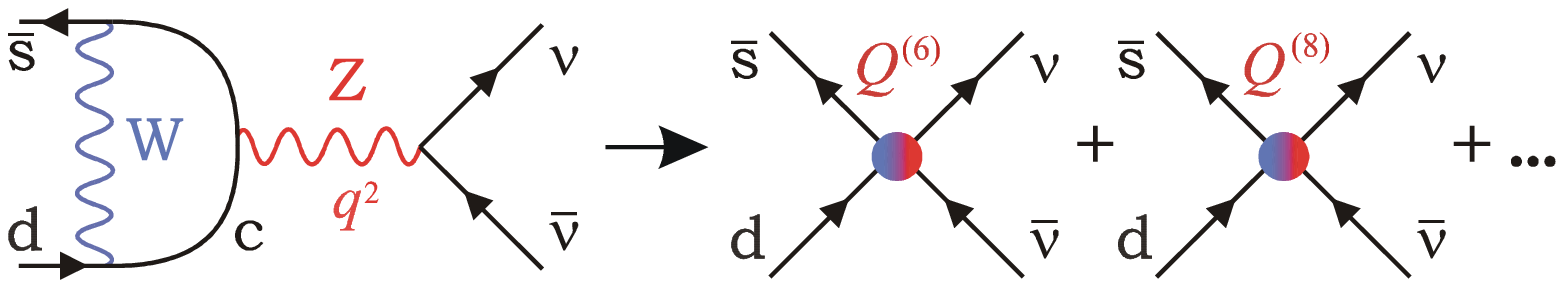}%
}%
$\;%
\begin{array}
[c]{c}%
\text{c-quark contribution }\\
\text{as a tower of local }\\
\text{interactions in }q^{n}/m_{c}^{n}%
\end{array}
$\\%
\raisebox{-0.9219in}{\includegraphics[
height=1.5618in,
width=4.6233in
]%
{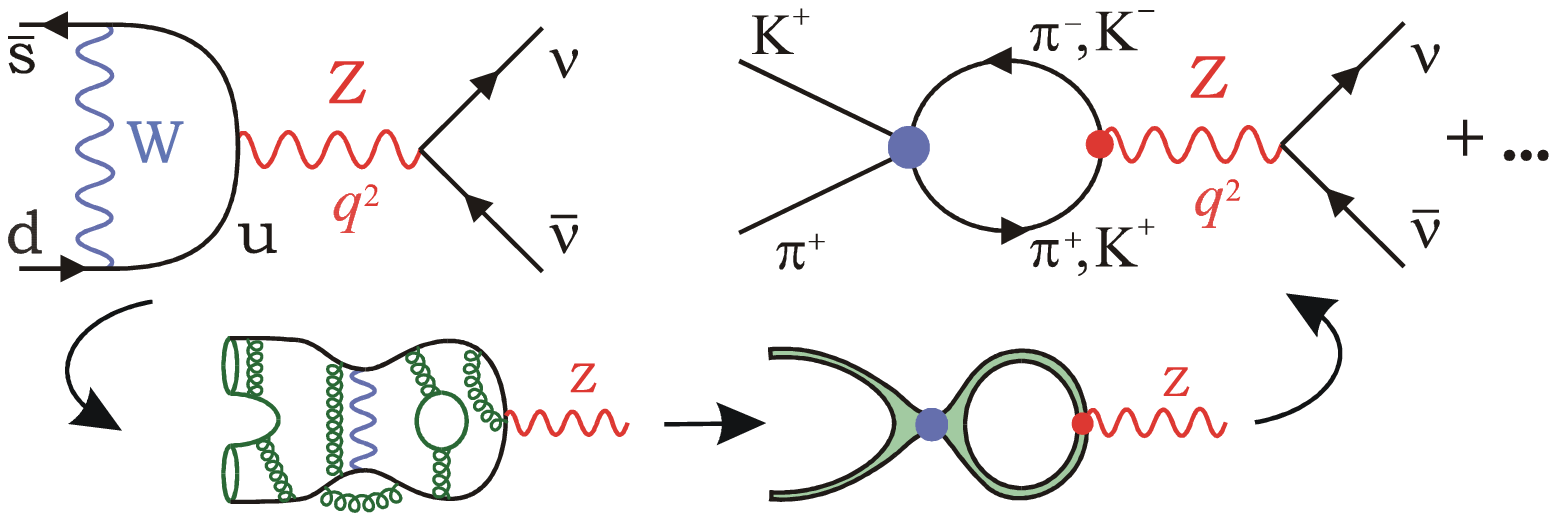}%
}%
$\;%
\begin{array}
[c]{c}%
\text{u-quark contribution }\\
\text{always non-local.}\\
\text{Meson loops dealt }\\
\text{with in ChPT.}%
\end{array}
$\\
\multicolumn{1}{c}{Figure 6: Schematic OPE structure including c-quark
$Q_{c}^{(8)}$ operators and LD meson loops.}%
\end{tabular}
\]
(similar expansions of the $W^{\pm}$ box are also considered). For dim. 8
operators $Q_{c}^{(8)}$ like $(\bar{s}d)_{V-A}\partial^{2}(\nu\bar{\nu}%
)_{V-A}$, we have confirmed by an OPE at the charm scale the basis found
in\thinspace\cite{Falk}. However, to estimate their matrix elements
$\langle\pi^{+}|Q_{c}^{(8)}|K^{+}\rangle$, we performed an (approximate)
matching with ChPT. Symbolically, if QCD is turned off, $Q_{c}^{(8)}$ and
$Q_{u}^{(8)}$ scale as
\begin{equation}
\lambda_{c}\frac{q^{2}}{M_{W}^{2}}\left(  \log x_{c}-\log x_{u}\right)
\rightarrow\lambda_{c}\frac{q^{2}}{M_{W}^{2}}\left(  \log\frac{m_{c}}{\mu
_{IR}}-\log\frac{m_{u}}{\mu_{UV}}\right)  \label{Eq12}%
\end{equation}
with $\mu_{IR}\equiv\mu_{UV}$ and $\lambda_{c}=-\lambda_{u}$. If this naive
picture survives to hadronization, there will be an exact cancellation of the
scale dependences from the Chiral loop UV divergences and the c-quark IR one.
Even if an exact matching cannot be expected (the ChPT amplitude exhibits the
$\Delta I=1/2$ enhancement), current estimates can be much improved.

For $u$-quark effects, the amplitude is computed at one-loop in ChPT. An
important improvement over previous analyses\thinspace\cite{LD} was to include
in the $\mathcal{O}(G_{F}^{2}p^{2})$ operator basis all the effective
interactions arising from the integration of heavy modes, in particular the
local $Z^{\mu}(\bar{s}d)_{V-A}$ non-gauge invariant one. Alternatively, one
can enforce GIM mechanism on the direct transition $K_{L}\rightarrow Z$,
leading to the $\Delta S=1$ ChPT Lagrangian\thinspace\cite{OurWorkNu}%
\begin{equation}
L_{\Delta S=1}^{(2)}=G_{8}F_{\pi}^{4}\left\{  \left\langle \lambda_{6}L_{\mu
}L^{\mu}\right\rangle -2i\frac{g}{\cos\theta_{W}}\left\langle \lambda
_{6}L_{\mu}T_{3}\right\rangle Z^{\mu}\right\}  \label{Eq13}%
\end{equation}
Then, contrary to earlier estimation, $K^{+}\rightarrow\pi^{+}Z^{\ast}$ is not
vanishing at tree-level, and does not depend on the singlet part of the Z
current. At one-loop, a divergence structure that matches the short-distance
$c$-quark one is found, and estimating $\langle\pi^{+}|Q_{c}^{(8)}%
|K^{+}\rangle$ is possible.

Numerically, the combined effect of $c$-quark dim. 8 operators and of
non-local long-distance $u$-quark loops can be written as $X\left(
x_{c}\right)  \rightarrow X\left(  x_{c}\right)  +\delta X\left(
x_{c}\right)  $ with $X\left(  x_{c}\right)  $ given earlier, and $\delta
X\left(  x_{c}\right)  =\lambda^{4}\left(  0.04\pm0.02\right)  $. This amounts
to an enhancement of about $6\%$ for the total rate, and is completely
dominated by the long-distance Z-penguin.

\section*{Conclusion}

Long-distance effects in rare $K$ decays are now under control. For
$K_{L}\rightarrow\pi^{0}\ell^{+}\ell^{-}$, the theoretical analysis of
\cite{DEIP} combined with recent measurements by NA48\thinspace\cite{NA48as}
permits the estimation of the indirect CPV contributions. Interference with
direct-CPV violation has been argued to be positive by two independent
groups\thinspace\cite{BDI,FGD}. For pure long-distance CP-conserving
contributions, the tensor $2^{++}$ one was estimated by \cite{BDI} from the
photon spectrum in $K_{L}\rightarrow\pi^{0}\gamma\gamma$ and is negligible.
For the scalar one, CPC-0$^{++}$, we found\thinspace\cite{OurWorkMu} that a
reliable estimate is possible thanks to the dynamical stability of the ratio
$\mathcal{B}\left(  K_{L}\rightarrow\pi^{0}\ell^{+}\ell^{-}\right)
/\mathcal{B}\left(  K_{L}\rightarrow\pi^{0}\gamma\gamma\right)  $. Obviously,
extensive use is made of experimental inputs in these analyses. In particular,
the main source of uncertainties on $\mathcal{B}\left(  K_{L}\rightarrow
\pi^{0}\ell^{+}\ell^{-}\right)  $ at present is in the parameter $a_{S}$ (see
Fig.5), so any improvement in the measurement of $\mathcal{B}\left(
K_{S}\rightarrow\pi^{0}\ell^{+}\ell^{-}\right)  $ would be much welcomed.

Concerning $K^{+}\rightarrow\pi^{+}\nu\bar{\nu}$, two subleading sources of
potentially large theoretical uncertainties were analyzed and brought under
control\thinspace\cite{OurWorkNu}, namely dimension eight $c$-quark operators,
and long-distance $u$-quark loops. Taken together, they amount to an increase
of the BR by $6\%$. Theoretical uncertainties are then dominated by the
$c$-quark dimension six OPE at NLO, and a NNLO analysis would presumably lead
to a theoretical error of less than 5\%. The ability of $K^{+}\rightarrow
\pi^{+}\nu\bar{\nu}$ in constraining the Standard Model is therefore lying on
strong theoretical grounds.\smallskip\newline \textbf{Acknowledgments:} This
work has been supported by IHP-RTN, EC contract No. HPRN-CT-2002-00311 (EURIDICE).


\section*{References}

\begin{thebibliography}{9}                                                                                                %

\bibitem {KTeV}A.~Alavi-Harati \textit{et al. }[KTeV],
Phys.\ Rev.\ \textbf{D61} (2000) 072006; Phys.\ Rev.\ Lett.\ \textbf{84}
(2000) 5279; Phys. Rev. Lett. \textbf{93} (2004) 021805.

\bibitem {E787E949}S.~Adler \textit{et al. }[E787], Phys. Rev. Lett.
\textbf{88} (2002) 041803; V.~V.~Anisimovsky \textit{et al. }[E949], Phys.
Rev. Lett. \textbf{93} (2004) 031801.

\bibitem {BLMM}A.J.~Buras, M.E.~Lautenbacher, M.~Misiak and M.~M\"{u}nz, Nucl.
Phys. \textbf{B423} (1994) 349; G. Buchalla, A.J. Buras and M.E. Lautenbacher,
Rev. Mod. Phys. \textbf{68} (1996) 1125; A.~J.~Buras, F.~Schwab and S.~Uhlig,
\textit{hep-ph/0405132}.

\bibitem {NA48as}J.R.~Batley et al. [NA48], Phys. Lett. \textbf{B576} (2003)
43; Phys. Lett. \textbf{B599} (2004) 197.

\bibitem {KTeV_KLpgg}A.~Alavi-Harati et al. [KTeV], Phys. Rev. Lett.
\textbf{83} (1999) 917.

\bibitem {NA48_KLpgg}A.~Lai et al. [NA48], Phys. Lett. \textbf{B536} (2002) 229.

\bibitem {OurWorkMu}G.~Isidori, C.~Smith and R.~Unterdorfer, Eur. Phys. J.
\textbf{C36} (2004) 57.

\bibitem {BDI}G.~Buchalla, G.~D'Ambrosio and G.~Isidori, Nucl. Phys.
\textbf{B672} (2003) 387.

\bibitem {DEIP}G.~D'Ambrosio, G.~Ecker, G.~Isidori and J.~Portol\'{e}s, JHEP
\textbf{08} (1998) 004.

\bibitem {FGD}S.~Friot, D.~Greynat, E. de Rafael, Phys. Lett. \textbf{B595}
(2004) 301.

\bibitem {BFRS}A.~J.~Buras, R.~Fleischer, S.~Recksiegel and F.~Schwab, Nucl.
Phys. \textbf{B697} (2004) 133.

\bibitem {OurWorkNu}G.~Isidori, F.~Mescia and C.~Smith, to appear in Nucl.
Phys. \textbf{B}, \textit{hep-ph/0503107}.

\bibitem {Falk}A.~F.~Falk, A.~Lewandowski and A.~A.~Petrov,
Phys.\ Lett.\ \textbf{B505} (2001) 107.

\bibitem {LD}D.~Rein and L.~M.~Sehgal, Phys.\ Rev.\ \textbf{D39} (1989) 3325;
J.~S.~Hagelin and L.~S.~Littenberg, Prog.\ Part.\ Nucl.\ Phys.\ \textbf{23}
(1989) 1; M.~Lu and M.~B.~Wise, Phys.\ Lett.\ \textbf{B324} (1994) 461.
\end{thebibliography}
\end{document}